\documentclass[prb,aps,twocolumn,showpacs,superscriptaddress]{revtex4-1}

\usepackage[dvips]{color,graphicx}
\DeclareGraphicsExtensions{.eps, .jpg, .png}

\usepackage{graphicx}
\usepackage{dcolumn}
\newcolumntype{d}{D{.}{.}{2}}
\usepackage{bm}
\usepackage{float}
\usepackage{color}
\usepackage{amsmath}
\usepackage{amssymb}

\usepackage{multirow}

\usepackage{pstricks}

\begin{document}

\title{Prediction of pressure-induced stabilization of noble-gas-atom compounds
with alkali oxides and alkali sulfides}

\author{Hao Gao} \affiliation{National Laboratory of Solid State
  Microstructures, School of Physics
  and Collaborative Innovation Center of Advanced Microstructures,
  Nanjing University, Nanjing 210093, China}

\author{Jian Sun} \email{jiansun@nju.edu.cn} \affiliation{National Laboratory of Solid State
  Microstructures, School of Physics
  and Collaborative Innovation Center of Advanced Microstructures,
  Nanjing University, Nanjing 210093, China}

\author{Chris J.\ Pickard} \affiliation{Department of Materials
  Science \& Metallurgy, University of Cambridge, 27 Charles Babbage
  Road, Cambridge CB3 0FS, UK} \affiliation{Advanced Institute for
  Materials Research, Tohoku University, 2-1-1 Katahira, Aoba, Sendai,
  980-8577, Japan}

\author{Richard J.\ Needs}
\affiliation{Theory of Condensed Matter Group, Cavendish Laboratory, J
  J Thomson Avenue, Cambridge CB3 0HE, UK}

\date{\today}

\begin{abstract}

  The cubic antifluorite structure comprises a face-centered cubic
  sublattice of anions with cations on the tetrahedral sites. The
  voids in the antifluorite structure that are crucial for
  superionicity in Li$_2$O might also act as atomic traps. Trapping of
  guest atoms and small molecules within voids of a host structure
  leads to the formation of what are known as clathrate compounds.
  Here we investigate the possibility of trapping helium or larger
  neon guest atoms under pressure within alkali metal oxide and
  sulfide structures. We find stable helium and neon-bearing compounds
  at very low pressures.  These structures are stabilized by a
  reduction in volume from incorporation of helium or neon atoms
  within the antifluorite structure. We predict that NeCs$_2$S could
  be stable at ambient pressure.  Our study suggests a novel class of
  alkali oxide and sulfide materials incorporating noble gas atoms
  that might potentially be useful for gas storage.

\end{abstract}

\maketitle

\section{Introduction}
\label{sec:intro}

\begin{figure}
  \includegraphics[width=0.44\textwidth]{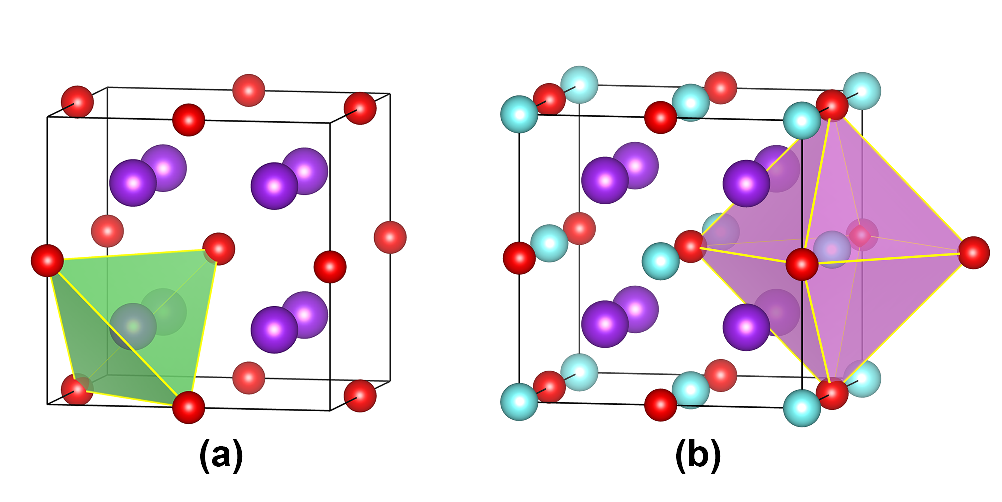}
  \caption{Crystal structure of antifluorite alkali metal oxide
  with and without helium atoms. (a) The antifluorite structure of
    Na$_2$O. The Na atoms are shown in purple and oxygen atoms in red.
    The Na atoms occupy tetrahedral positions, and one of the Na atoms
    is shown at the center of its green tetrahedron.  (b) He atoms
    (light blue) are included within the octahedral voids, and one of
    the octahedrons is highlighted in purple with a He atom visible at
    its center.}
  \label{fig:antifluorite_with_atoms_voids}
\end{figure}

Alkali metal oxides are used in catalysing reactions
\cite{Tsuji_1991}, capture and storage of CO$_2$ \cite{Wang_2011}, gas
detectors, and for lowering the work function of photo-cathodes
\cite{Martinez_2008}.  Lithium oxide (Li$_2$O) is a key battery
material, and it is also used within the nuclear industry.  At low
pressures alkali oxides
\cite{Mikajlo_Li2O_2002,Mikajlo_Na2O_2003,Mikajlo_K2O_2003,Kunc_2005_Li2O,Cancarevic_alkali-metal_oxides_2006,Lazicki_Li2O_2006,Islam_Li2O_2006,Eithiraj_alkali-metal_oxides_2007,Moakafi_alkali-metal_oxides_2008,Long_Li2O_2013}
and sulfides
\cite{Sommer_1977_Cs2S,Grzechnik_Li2S_2000,Vegas_Na2S_2001,Vegas_K2S_2002,Santamaria-Perez_Rb2S_2011,Santamaria-Perez_Cs2S_2011}
normally crystallize in the antifluorite structure of space group
$Fm\bar{3}m$, which contains octahedral voids, see Fig.\
\ref{fig:antifluorite_with_atoms_voids}.  Li$_2$O exhibits superionic
conductivity at temperatures above 1200 K in which diffusing Li$^{+}$
ions carry electrical current by hopping from one void to another,
while the oxygen atoms remain within a rigid framework
\cite{Varshney_superionic_LiO2,Hull_Superionics_2004}.  The voids in
the antifluorite structure that are crucial for superionicity in
Li$_2$O might also act as atomic traps.  Trapping of guest atoms and
small molecules within voids of a host structure can lead to the
formation of what are known as clathrate compounds
\cite{Mao_clathrates}.  Clathrates involving many different guest
species have been observed, such as clathrate hydrates with inclusion
of helium (He), neon (Ne), argon (Ar)
\cite{Londono_1988_helium,Barrer_1962_neon,teeratchanan_computational_2015},
or other noble gas atoms. Trapped noble gas atoms have also been found
in alkali metals \cite{dong_stable_2017} and oxides
\cite{sans_ordered_2016,liu_reactivity_2018} under pressure. Such
compounds have been reviewed by Grochala \cite{Grochala_noble_2007}.

Here we investigate the possibility of trapping He or larger Ne guest
atoms under pressure within alkali metal oxide and sulfide structures.
By studying different noble gas atoms, alkali oxides and sulfides, we
are able to assess the effects of changing both the host and guest
species.  At zero pressure the thermodynamic driving force for
including He or Ne atoms within the alkali oxides or sulfides is
small.  The inertness of He and Ne suggests that they might be
included within structures containing voids without substantial
changes in the host structure.  Structures containing voids are
unlikely to be stable at high pressures, and it is more likely that
the thermodynamically stable state at high pressures of, say, a
mixture of He and Li$_2$O, will consist of hexagonal-close-packed
(hcp) He and a dense high-pressure phase of Li$_2$O.  The energetic
stability of structures consisting of He or Ne atoms trapped in the
antifluorite structures of alkali oxides and sulfides therefore rests
on a delicate balance of competing effects.

The ambient and high pressure structures of alkali metal sulfides have
been studied extensively, and good agreement between experiment and
the results of first-principles density-functional-theory (DFT)
calculations has been found
\cite{Cancarevic_alkali-metal_oxides_2006}.  There is also good
agreement between experiment and DFT results for alkali metal oxides
under ambient conditions, although their high-pressure behavior has
not been studied in depth experimentally, except for Li$_2$O
\cite{Kunc_2005_Li2O,Lazicki_Li2O_2006}.

A transition from the antifluorite to an orthorhombic anticotunnite
structure of $Pnma$ symmetry has been observed in Li$_2$O at pressures
around 50 GPa \cite{Kunc_2005_Li2O,Lazicki_Li2O_2006}.  Transitions
from the antifluorite to anticotunnite structures have also been
reported in Li$_2$S at 12 GPa \cite{Grzechnik_Li2S_2000}, Na$_2$S at 7
GPa \cite{Vegas_Na2S_2001} and in Rb$_2$S below 0.7 GPa
\cite{Santamaria-Perez_Rb2S_2011}.  Interestingly, the anticotunnite
structure has not been observed in K$_2$S, and a distorted
Ni$_2$In-type structure of space group $P6_3/mmc$ has instead been
found at 6 GPa \cite{Vegas_K2S_2002}.  A transition from the
antifluorite to anticotunnite structure has been observed in Rb$_2$S
below 0.7 GPa, followed by a transition to a Ni$_2$In-type structure
at 2.6 GPa \cite{Santamaria-Perez_Rb2S_2011}.  Cs$_2$S has been
reported to crystallize in the anticotunnite structure
\cite{Sommer_1977_Cs2S} under ambient conditions, and to transform
into a distorted Ni$_2$In-type structure by about 5 GPa
\cite{Santamaria-Perez_Cs2S_2011}.

\section{Methods}
\label{sec:method}

\renewcommand\arraystretch{1.5}
\begin{table}
  \begin{ruledtabular}
    \caption{Lattice constant ({\AA}) of K$_2$O with different
      exchange-correlation functionals and K pseudopotentials.}
    \begin{tabular}{c c c}%
      \multirow{2}*{Exchange-Correlation}
      & \multicolumn{2}{c}{Valence electrons}\\\cline{2-3}
      {Functionals} & $3p^64s^1$  &   $3s^23p^64s^1$\\
      optB88-vdW \cite{klimes_chemical_2010}  & 6.435 & 6.382 \\
      optB86b-vdW \cite{klimes_van_2011}  & 6.432 & 6.373 \\
      vdW-DF2 \cite{lee_higher-accuracy_2010}    & 6.557 & 6.518 \\
      DFT-D2 \cite{grimme_semiempirical_2006}     & 6.452 & 6.405 \\
      DFT-D3 \cite{grimme_consistent_2010}     & 6.494 & 6.436 \\
      LDA         & 6.272 & 6.488 \\
      PBE         & 6.537 & 6.487 \\
      PBEsol \cite{perdew_restoring_2008}     & 6.413 & 6.347 \\
      \hline
      Experiment(317K) \cite{touzain_systemes_1970}  & \multicolumn{2}{c}{6.436}   \\
    \end{tabular}
    \label{table:lattice_K2O}
  \end{ruledtabular}
\end{table}

\begin{table}
  \begin{ruledtabular}
    \caption{Lattice constant ({\AA}) of HCP He (18MPa)}
    \begin{tabular}{c c c}%
      \multirow{2}*{Exchange-Correlation}
      & \multicolumn{2}{c}{Lattice constant}\\\cline{2-3}
      {Functionals} & a  &   c\\
      optB88-vdW  & 2.885 & 4.756 \\
      DFT-D3      & 2.712 & 4.444 \\
      LDA         & 2.426 & 3.988 \\
      PBE         & 2.864 & 4.699 \\
      PBEsol      & 3.016 & 4.948 \\
      \hline
      Experiment(3.3K) \cite{schuch_structure_1958}   &  3.460 & 5.600  \\
    \end{tabular}
    \label{table:lattice_He}
  \end{ruledtabular}
\end{table}

\begin{table}
  \begin{ruledtabular}
    \caption{Lattice constant ({\AA}) of FCC Ne}
    \begin{tabular}{c c c}%
      \multirow{2}*{Exchange-Correlation}
      & \multicolumn{2}{c}{Pressures}\\\cline{2-3}
      {Functionals} & 0  &   10GPa\\
      optB88-vdW  & 4.264 & 3.560 \\
      DFT-D3      & 4.345 & 3.601 \\
      LDA         & 3.852 & 3.442 \\
      PBE         & 4.510 & 3.619 \\
      PBEsol      & 4.560 & 3.563 \\
      \hline
      Experiment   &  4.42(4.2K)\cite{henshaw_atomic_1958} & 3.567(300K)\cite{hemley_x-ray_1989}   \\
    \end{tabular}
    \label{table:lattice_Ne}
  \end{ruledtabular}
\end{table}

We used the \textit{ab initio} random structure searching (AIRSS)
method and the \textsc{Castep} plane-wave DFT code
\cite{ClarkSPHPRP05} to find stable structures of alkali oxides and
sulfides containing He or Ne atoms.  In the AIRSS approach randomly
chosen structures are relaxed to local minima in the enthalpy
\cite{PickardN06_silane,Airss_review,needs_perspective_2016}.  AIRSS
has been successfully applied to many systems, including alkali metals
\cite{Pickard_lithium_2009}, and various oxides such as carbon
monoxide \cite{Sun_CO2_2011, Xia_CO_2017}.  The electronic structures
and enthalpies were calculated using the \textsc{Castep} code. Very
similar results were obtained with the projector augmented wave (PAW)
method implemented in the Vienna \textit{ab initio} simulation package
(VASP)~\cite{vasp}.

We performed AIRSS calculations for a variety of Na/O/He
stoichiometries at a pressure of 15 GPa.  The most stable Na/O
compound at 15 GPa was found to be Na$_2$O in the antifluorite
structure, as expected, and we found the anticotunnite structure and a
Ni$_2$In-type structure of space group $P6_3/mmc$.  We also found a
structure of chemical composition HeNa$_2$O in which Na$_2$O adopts
the antifluorite structure and the He atoms sit at the octahedral
voids, so that the $Fm\bar{3}m$ symmetry of the antifluorite structure
is maintained.  The HeNa$_2$O structure was found to be slightly more
stable than antifluorite Na$_2$O at 15 GPa.  We proceeded to
investigate the possibility of stable antifluorite structures of
alkali oxides and sulfides with He or Ne atoms placed at the
octahedral voids.  We then widened our study to include other
competitive structures of the alkali oxides and sulfides. As shown in
TABLE \uppercase\expandafter{\romannumeral1},
\uppercase\expandafter{\romannumeral2} and
\uppercase\expandafter{\romannumeral3}, we have calculated the cell
parameters of K$_2$O, He and Ne with several different
exchange-correlation functionals and pseudopotentials, and compared
them with experimental data.  The results demonstrate the necessity of
using pseudopotentials with nine valence electrons for alkali metal
elements.  The lattice parameters of solid He calculated by DFT is
about $10\%$ lower than in the experimental data. The deviations are
mainly attributed to nuclear quantum effects and zero-point
energy. Other \textit{ab initio} methods, such as Diffusion quantum
Monte Carlo (DMC) can be performed to obtain accurate results in
excellent agreement with experiment
\cite{cazorla_zero-temperature_2008, needs_continuum_2010}.
Grimme's dispersion corrected
(DFT-D) methods
\cite{grimme_semiempirical_2006,grimme_consistent_2010} perform very
well in this test; however, the parameters were fitted to experimental
data at ambient pressure.  Therefore they may not be suitable for
calculations at high pressures.  The vdW-DF corrections of Lundqvist
\textit{et al.}\ \cite{vdw_Lundqvist_2004} have a non-local term that
is dependent on the electron density, which should be more appropriate
than DFT-D methods at high pressure.  Therefore we have used the
optB88-vdW functional \cite{klimes_chemical_2010} and
pseudopotentials with 9 valence electrons for alkali metals in our
plane-wave calculations.  A basis set energy cutoff of 800 eV was
used, except for Na compounds, in which a higher cutoff energy of 980
eV was used. The Brillouin zone was sampled using a $k$-point grid
spacing of $2\pi\times$0.02~\AA$^{-1}$ for the final converged results
reported in this paper. We have also performed calculations using
Heyd-Scuseria-Ernzerhof (HSE06) hybrid functionals
\cite{heyd_hybrid_2003} which provide accurate electronic band gaps
for Na$_2$O and HeNa$_2$O when using pseudopotentials with 7 valence
electrons.

\section{Results and Discussion}
\label{sec:res}

\begin{figure*}
\centering
		\includegraphics[width=0.82\textwidth]{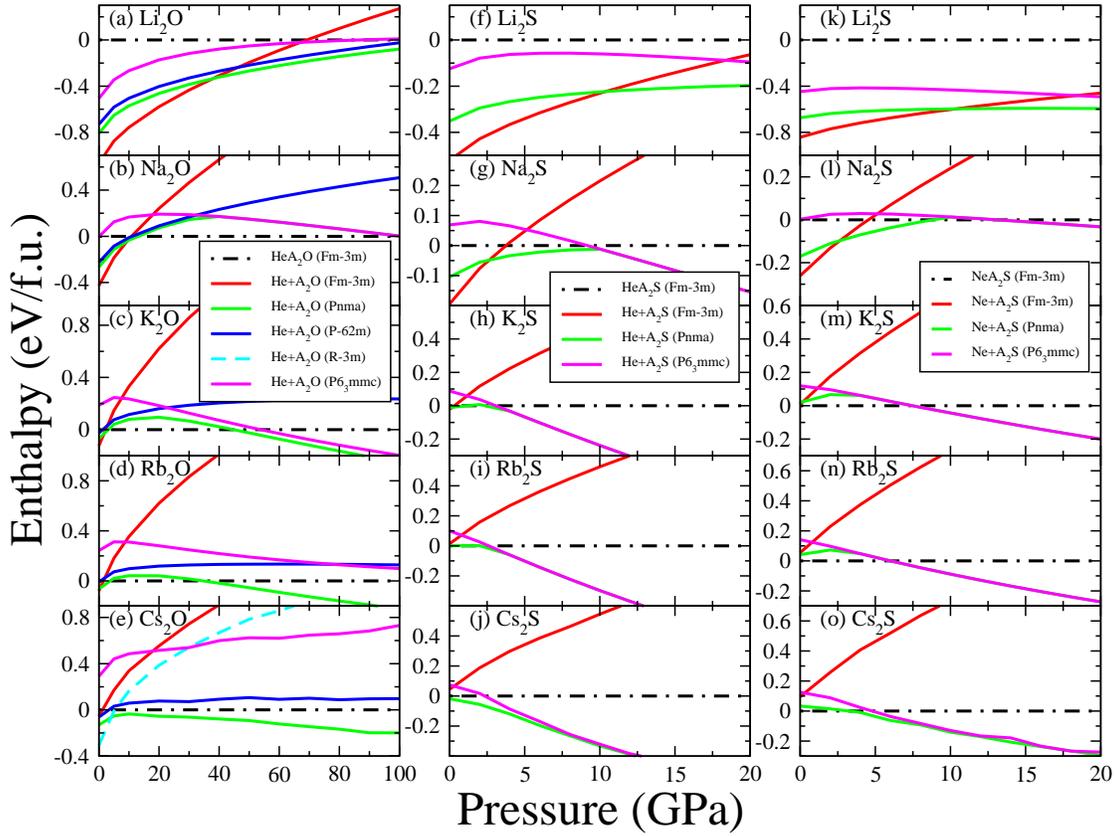}
                \caption{Enthalpy-pressure relations for noble gas
                  compounds.  (a--e) He in alkali metal oxides, (f--j) He in
                  alkali metal sulfides, and (k--o) Ne in alkali metal
                  sulfides.  The enthalpies are given with respect to
                  those of the alkali oxides/sulfides with included
                  noble gas atoms.}
\label{fig:enthalpy_pressure}
\end{figure*}

As well as the antifluorite and anticotunnite structures, we
considered the CdCl$_2$ ($R\bar{3}m$), $\gamma$US$_2$ ($P\bar{6}2m$),
and CaCl$_2$ ($Pnnm$) structures of the oxides, and $Pnnm$, $Pmma$ and
the $P6_3/mmc$ structures of the alkali metal sulfides.  We found that
the enthalpy-pressure curves of the $Pmma$ phase reported for K$_2$S
in Ref.\ \onlinecite{Vegas_K2S_2002} were very close to those of the
$P6_3/mmc$ structure for the sulfides from Li$_2$S to Cs$_2$S at the
pressures studied, and therefore we have not reported results for the
$Pmma$ structure.  The energetically competitive structures were
relaxed, and their relative enthalpies are reported in Fig.\
\ref{fig:enthalpy_pressure}.  Interestingly, we found that the
transition from the antifluorite to anticotunnite structure in Na$_2$O
occurs at fairly low pressures, but at higher pressures a transition
back to antifluorite occurs with included He atoms.
Further details of the phase transition pressures are reported in the
Supplemental Material \cite{Supplemental}.

Consider a static lattice calculation of the total enthalpy $H=U+pV$,
where $U$ is the total internal energy, $p$ is the pressure and $V$ is
the volume.  Taking Na$_2$O + He as an example, the reduction in the
total volume on forming HeNa$_2$O($Fm\bar{3}m$) at 20 GPa corresponds
to about {\small 1/4} of the volume of a He atom in hcp He at the same
pressure.  We find that the formation of HeNa$_2$O($Fm\bar{3}m$) + hcp
He from He+Na$_2$O($Fm\bar{3}m$) increases the total internal energy
by 244 meV per formula unit, but the $pV$ term decreases by 486 meV
per formula unit.  This shows that although the interactions between
the He atoms and the host antifluorite Na$_2$O are not energetically
favorable, the reduction in volume from inclusion of the He atoms
lowers the overall enthalpy.  The pressure derivative of the enthalpy
is equal to the volume ($dH/dp = V$), so that the slopes of the curves
in Fig.\ \ref{fig:enthalpy_pressure}(b) give the volumes relative to the
reference HeNa$_2$O($Fm\bar{3}m$) phase (denoted by the dashed-dotted
line). The most stable form at low pressures is Na$_2$O($Fm\bar{3}m$)
+ hcp He, which transforms into Na$_2$O($Pnma$) + hcp He at 9.3 GPa.
At 13 GPa a transformation occurs to HeNa$_2$O($Fm\bar{3}m$), which
is stable up to just above 100 GPa.  We observe from the slopes of the
enthalpy-pressure curves in Fig.\ \ref{fig:enthalpy_pressure}(b) that at
13 GPa the volume of HeNa$_2$O($Fm\bar{3}m$) is smaller than that of
the other Na$_2$O + He phases.  The thermodynamic driving force for
the initial incorporation of He atoms is therefore the associated
reduction in total volume.  A maximum in the enthalpy reduction from
He inclusion in Na$_2$O occurs at about 40 GPa.  At higher pressures
the stability of the inclusion compound is reduced because the voids
in the compressed antifluorite structure are no longer large enough to
accommodate the He atoms, which results in an increase in the internal
energy, and HeNa$_2$O($Fm\bar{3}m$) eventually becomes unstable to
formation of the $P6_3/mmc$ structure just above 100 GPa, see Fig.\
\ref{fig:enthalpy_pressure}.

\begin{figure*}
  \centering
  \includegraphics[width=0.82\textwidth]{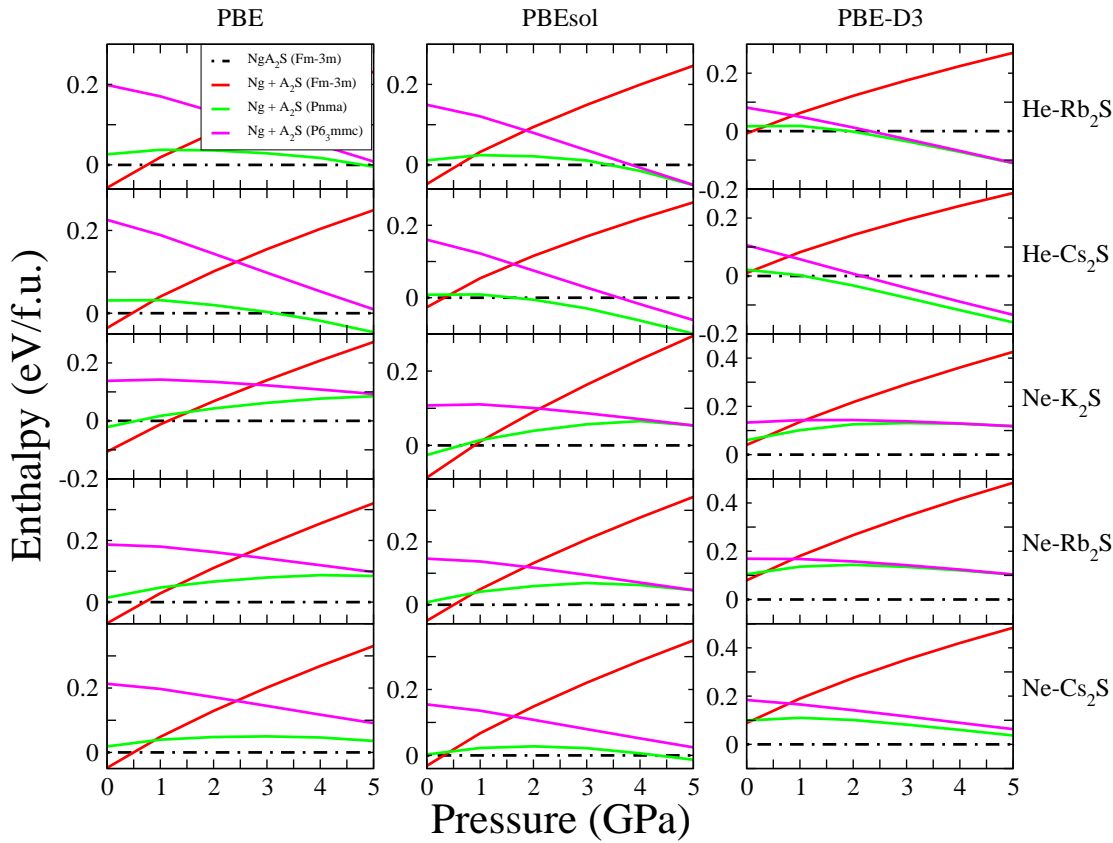}
  \caption{Enthalpy-pressure relations for noble gas compounds
    calculated with the PBE, PBEsol and DFT-D3 functionals.}
  \label{fig:enthalpy_pressure_low}
\end{figure*}

The transition pressures for noble gas inclusion are lower for the
sulfides than the oxides because the larger size of the S atom leads
to larger octahedral voids in the sulfide structures.  There is an
upper pressure limit to the stability of each of the noble gas
compounds discussed here, which depends on the sizes and chemical
identities of the host and guest species.  The larger stabilization
enthalpies of Ne atoms in alkali sulfides compared with He is due to
the better matching of the size of the Ne atom with that of the voids
in the antifluorite sulfide structures.  More $pV$ energy is gained by
including a Ne atom within the antifluorite structure rather than a
smaller He atom.  If the void is large enough the additional strain
energy from including the larger Ne atom is small, and it is
energetically favorable to include the Ne atom.  In other words, the
stabilization enthalpy is largest at pressures for which the sizes of
the voids and included alkali atoms are well matched.



\begin{figure}
  \includegraphics[width=0.42\textwidth]{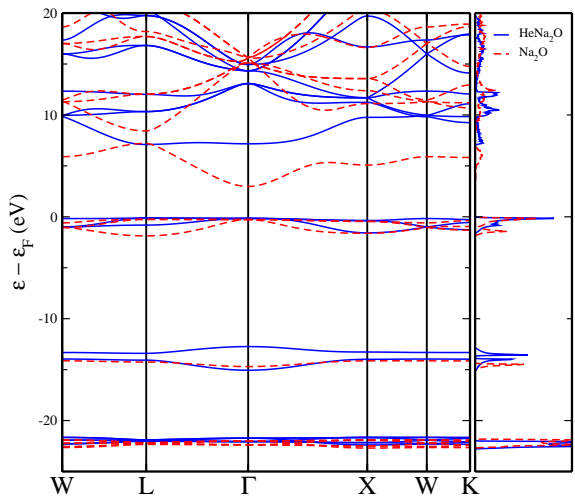}
  \caption{Electronic Band structure and Density of states of Na$_2$O
    (red dashed line) and HeNa$_2$O (blue solid line) at 20 GPa.  The
    zero of energy is at the valence band maximum.  The He $1s$ band
    lies at about -13.5 eV.}
  \label{fig:bandstructure_HeNa2O-Na2O}
\end{figure}

We find ranges of stability for antifluorite HeK$_2$O and HeRb$_2$O,
as well as HeNa$_2$O, but HeLi$_2$O and HeCs$_2$O are
thermodynamically unstable at the pressures studied, see Fig.\
\ref{fig:enthalpy_pressure}(a--e).  The stability ranges are 13--105 GPa
(HeNa$_2$O), 3--45 GPa (HeK$_2$O), and 3.7--34.5 GPa (HeRb$_2$O).
Replacing the O atoms by larger S atoms we find that HeLi$_2$S,
HeNa$_2$S and HeCs$_2$S are not thermodynamically stable, but that HeK$_2$S and
HeRb$_2$S have small regions of stability at low pressures with small
stabilization enthalpies, and HeRb$_2$S is predicted to be stable at
ambient pressure.  The stability ranges are 1.3--2.4 GPa (HeK$_2$S)
and 0--2 GPa (HeRb$_2$S), see Fig.\ \ref{fig:enthalpy_pressure}(f--j).  The
inclusion of He atoms within the alkali sulfides is therefore not
particularly favorable.  Replacing the He atoms by larger Ne atoms, we
find that NeNa$_2$S, NeK$_2$S, NeRb$_2$S, and NeCs$_2$S are predicted to be
stable under ambient or high pressure, but NeLi$_2$S is
unstable up to 20 GPa.  The stability ranges are also larger: 8.8--13 GPa (NeNa$_2$S),
0--7.5 GPa (NeK$_2$S), 0--6 GPa (NeRb$_2$S), and 0--3.3 Ga (NeCs$_2$S), see
Fig.\ \ref{fig:enthalpy_pressure}(k--o).  The maximum stabilization energies
for the helium alkali oxides, helium alkali sulfides, and neon alkali
sulfides are: 172 meV (HeNa$_2$O at 40 GPa); 8 meV (HeK$_2$S at 2
GPa); and 72 meV (NeRb$_2$S at 2 GPa).

To validate the stability of the noble gas ternary compounds at low
pressures we calculated enthalpy-pressure relations in the range 0--5
GPa using the PBE, PBEsol \cite{perdew_restoring_2008} and DFT-D3
\cite{grimme_consistent_2010} functionals.  As can be seen in Fig.\
\ref{fig:enthalpy_pressure_low}, including the vdW interactions
reduces the formation pressures of noble gas compounds.  Different vdW
correction methods, such as optB88-vdW \cite{klimes_chemical_2010}
and DFT-D3 \cite{grimme_consistent_2010}, give similar results. We
predict that NeK$_2$S, NeRb$_2$S and NeCs$_2$S could exist at ambient
pressure.

Each of the materials investigated is an insulator.  The
bandstructures of the alkali oxides and sulfides consist of narrow and
rather widely spaced bands, as shown in Fig.\
\ref{fig:bandstructure_HeNa2O-Na2O} for Na$_2$O at 20 GPa.  The
orbitals at the valence band maximum arise mainly from the O $2p$
levels, which capture the valence $3s$ electrons from the Na atoms.
The bands at about -14.5 eV arise from the O $2s$ levels, and those at
-22.0 eV arise from the Na $2p$ levels.  Additional H atoms occupy a
band at about -13.5 eV which shows a small amount of hybridization
with the O $2s$ levels.  The Na $2s$ levels lie at about -50 eV (not
shown).

In K$_2$S at 2 GPa the O $2p$ levels capture the K $4s$ electrons,
while the O $2s$ levels lie at about -9.4 eV.  The K $3p$ levels lie
at about -12.6 eV.  When He atoms are included they occupy a $1s$ band
which is very close in energy to the K $3p$ levels, and there is
considerable hybridization between these orbitals, see the
Supplemental Material \cite{Supplemental}.  If Ne atoms are included
instead of He, we find that the Ne $2p$ levels sit at an energy of
about -10.0 eV, just below the O $2s$ levels.

The inclusion of noble gas atoms does not in general alter the oxygen,
sulfur or alkali-metal-derived bands very much, which is consistent
with the interaction between the noble gas atoms and the host
oxide/sulfide being fairly weak.  However, in some cases significant
hybridization between the alkali atom and O $2s$ levels can occur.
Bandstructures for other oxides and sulfides, with and without
included He and Ne atoms are shown in the Supplemental Material
\cite{Supplemental}.

Inclusion of He or Ne atoms within the alkali oxides and sulfides
increases their band gaps.  For example, at 2 GPa the band gap of
K$_2$S is increased by about 0.4 eV on inclusion of Ne atoms, and by
1.2 eV for He atoms.  The band gap of Na$_2$O at 20 GPa is increased
by 1.1 eV on including Ne atoms, but by 3.5 eV for He atoms.  In the
cases that we have studied, the band gaps are increased more by
including He atoms than Ne atoms.  We have investigated the origin of
the increased band gaps by performing calculations for the oxides and
sulfides at the larger volumes of the compounds that include noble gas
atoms.  We find that the increased gaps are mainly due to the presence
of the noble gas atoms, rather than the increase in volume.  This
arises because the nearly-free-electron states (parabolic band around
$\Gamma$ point) at the conduction band minimum have a significant
weight within the voids of the antifluorite structure, and the
presence of the noble gas atoms make the states more localized with
almost no dispersion.
The band gaps obtained using the PBE functional are a factor of about
two smaller than experimental values for Li$_2$O (8 eV
\cite{Ishii_1999_Li}), Na$_2$O (4.4--5.8 eV \cite{Rauch_1940_Na2O}),
and K$_2$O (4.0-5.4 eV \cite{Rauch_1940_Na2O}). The results from
quasiparticle bandstructure calculations within the $G_0W_0$
approximation \cite{Sommer_alkali-metal_oxides_2012} or
self-interaction corrections \cite{baumeier_electronic_2008} are in
quite good agreement with the available experimental data.
Underestimation of band gaps within standard DFT calculations is
well-known, but the changes in band gaps due to the introduction of
the noble gas atoms are likely to be significantly more accurate than
the gaps themselves. To confirm this conclusion, we have used a HSE06
hybrid functional, which has been successfully applied to Li-ion
battery electrode materials \cite{qi_lithium_2014} to calculate
accurate electronic band structures of Na$_2$O and HeNa$_2$O, as shown
in the Supplemental Material \cite{Supplemental}.  Compared with PBE
results, both band gaps increase by about 1.8 eV and the changes in
band gap caused by inclusion of He are almost same.

\begin{figure}[!hbp]
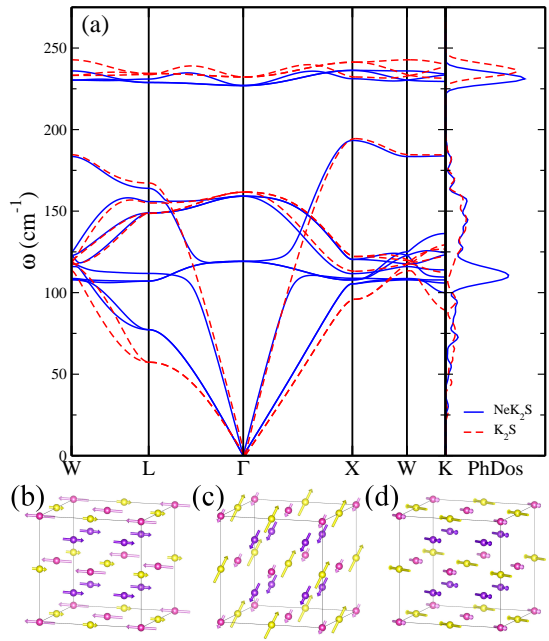

  \includegraphics[width=0.4\textwidth]{Fig5a-vdw-phono.eps}
  \hspace{0.5cm}\includegraphics[width=0.4\textwidth]{Fig5bcd-normal_mode.eps}

  \caption{(a) Phonon bandstructures and densities of states of K$_2$S
    (red dashed line) and NeK$_2$S (blue solid line) at 5 GPa.  The
    Ne-derived vibrations give Einstein-oscillator-like phonon bands
    at about 110 cm$^{-1}$ which show some hybridization with the
    K$_2$S-derived bands. (b) The localized modes obtained for Ne at
    the $\Gamma$ point. (c)(d) The localized modes with highest
    frequency at the $\Gamma$ point. The vectors in (b)(c)(d) show the
    polarization directions and amplitudes of the atoms. Ne, K, S
    atoms are colored magenta, purple and yellow respectively.}
  \label{fig:phonon_bandstructure_NeK2S-K2S}
\end{figure}

The phonon bandstructures of K$_2$S and NeK$_2$S at 5 GPa shown in
Fig.\ \ref{fig:phonon_bandstructure_NeK2S-K2S} indicate that the Ne
atoms give rise to Einstein-oscillator-like bands at about 110
cm$^{-1}$.  We have investigated the dynamical stability of the
$Fm\bar{3}m$ structures with incorporated noble gas atoms, finding
that the phonon bandstructures of HeK$_2$S and HeNa$_2$O have no
imaginary frequency modes down to zero pressure, and therefore they
might be quench recoverable under ambient conditions.

HeK$_2$S is, however, dynamically unstable at low pressures, but it is
stabilized by pressures of a few GPa, see Figs.\ 6 and 7 of the
Supplemental Material \cite{Supplemental}.  It is neccessary to
consider thermal expansion of crystals
\cite{wrobel_thermodynamic_2012}, and therefore we have corrected the
lattice constants and enthalpies of NeK$_2$S using the quasiharmonic
approximation (QHA).  The lattice parameter changes from 7.04\AA ~to
7.09\AA, with a minor difference of 0.7\%.  At the static lattice
level and a pressure of 5 GPa, NeK$_2$S is about 80 meV/fu more stable
than Ne+K$_2$S, but when quasiharmonic phonon enthalpies are added the
stability of NeK$_2$S is reduced to 54 meV/fu.  Similarly, at the
static lattice level and a pressure of 40 GPa, HeNa$_2$O is about 187
meV/fu more stable than He+Na$_2$O.  Including vibrational zero-point
energy reduces this to 175 meV/fu.  We have found strong evidence that
materials such as HeNa$_2$O and NeK$_2$S could be thermodynamically
stable under pressure, and that it might be possible to synthesize
them.  More information about the phonon bandstructures is provided in
the Supplemental Material \cite{Supplemental}.

The inclusion of noble gas atoms might be detected via an increase in
the lattice constant observable in x-ray diffraction, or in Raman or
infrared (IR) vibrational spectroscopy.  The antifluorite oxides and
sulfides, without noble gas atoms, have one Raman and one IR active
vibrational mode.  Inclusion of Ne atoms in K$_2$S at 5 GPa leads to
the appearance of an additional vibrational mode at around 110
cm$^{-1}$, as can be seen in Fig.\
\ref{fig:phonon_bandstructure_NeK2S-K2S}.  The Ne-derived normal modes
represent the relative motion between Ne atoms and K$_2$S crystal, as
shown in Fig.\ \ref{fig:phonon_bandstructure_NeK2S-K2S}(b), and the
vibrational amplitude of Ne is much higher than that of K and S.
Other localized high frequency modes are mainly derived from the S
atoms (Fig.\ \ref{fig:phonon_bandstructure_NeK2S-K2S}(c)(d)). The
vibrations are in directions along body diagonals or face diagonals.
The zone-center mode of NeK$_2$S at around 110 cm$^{-1}$ is IR active,
and the NeK$_2$S/K$_2$S modes around 150 cm$^{-1}$ are Raman active,
and those around 230 cm$^{-1}$ are IR active.  The calculated Raman
and IR intensities for NeK$_2$S/K$_2$S at 5 GPa are shown in the
Supplemental Material \cite{Supplemental}.  The appearance of a new IR
active vibrational mode of NeK$_2$S at around 110 cm$^{-1}$, the
shifts in the Raman and IR frequencies, and the changes in Raman
intensity could be used as signatures of the inclusion of Ne atoms.
Similar data for the inclusion of He within Na$_2$O at 40 GPa are
reported in the Supplemental Material \cite{Supplemental}.  As
condensed He or Ne are commonly used as the pressure transmission
medium in diamond-anvil-cell experiments \cite{Jayaraman_DAC_1983},
our results are also relevant to the question of whether these atoms
could enter alkali oxides and sulfides in diamond-anvil-cell
experiments.

\section{Conclusion}
\label{sec:end}

In summary, we have predicted the stability of compounds at high
pressures in which He or Ne atoms are included within alkali metal
oxides or sulfides.  In some cases structures including noble gas
atoms are predicted to become thermodynamically stable at pressures
below 1 GPa.  The chemical interactions between the host oxide or
sulfide and the guest noble gas atoms is weak, and the materials
discussed here may be described as inclusion compounds.  Our study
suggests a new class of metal/noble gas inclusion compounds that could
be synthesized under applied pressure, and might be useful for gas
storage. It seems likely that many more compounds containing noble gas
atoms could be stabilized by pressure.  We predict that some Ne
compounds, such as NeK$_2$S, NeRb$_2$S, and NeCs$_2$S might be stable
at ambient pressure.  We hope that our study may encourage attempts to
synthesize this class of compounds.

\begin{acknowledgments}
  The authors thank Tong Chen for the help on HSE06 calculations.
  J.S.\ gratefully acknowledges financial support from the MOST of
  China (Grant Nos.\ 2016YFA0300404, 2015CB921202), the National
  Natural Science Foundation of China (Grant Nos.\ 11574133 and
  11834006), the NSF of Jiangsu Province (Grant No.\ BK20150012),
  Special Program for Applied Research on Super Computation of the
  NSFC-Guangdong Joint Fund (the second phase) under Grant No.\
  U1501501, the Science Challenge Project (No.\ TZ2016001), and the
  Fundamental Research Funds for the Central Universities.  C.J.P.\
  and R.J.N.\ acknowledge financial support from the Engineering and
  Physical Sciences Research Council (EPSRC) of the U.K.\ under grants
  [EP/G007489/2] (C.J.P.) and [EP/P034616/1] (R.J.N.).  C.J.P.\ also
  acknowledges financial support from EPSRC and the Royal Society
  through a Royal Society Wolfson Research Merit award.  The
  calculations were carried out using supercomputers at Nanjing
  University, ``Tianhe-2'' at NSCC-Guangzhou and the CSD3 Peta4
  CPU/KNL machine in the University of Cambridge.
\end{acknowledgments}

%

\end{document}